\documentclass[12pt]{iopart}

\usepackage{graphicx}

\begin{document}

\title[]{Non-equilibrium quadratic measurement-feedback squeezing in a micromechanical resonator}

\author{Motoki Asano$^1$, Takuma Aihara$^2$, Tai Tsuchizawa$^2$, and Hiroshi Yamaguchi$^1$}

\address{$^1$NTT Basic Research Laboratories }
\address{$^2$NTT Device Technology Laboratories}
\ead{motoki.asano.gm@hco.ntt.co.jp}
\vspace{10pt}
\begin{indented}
\item[]
\end{indented}

\begin{abstract}
Measurement and feedback control of stochastic dynamics has been actively studied for not only stabilizing the system but also for generating additional entropy flows originating in the information flow in the feedback controller. In particular, a micromechanical system offers a great platform to investigate such non-equilibrium dynamics under measurement-feedback control owing to its precise controllability of small fluctuations. Although various types of measurement-feedback protocols have been demonstrated with linear observables (e.g., displacement and velocity), extending them to the nonlinear regime, i.e., utilizing nonlinear observables in both measurement and control, retains non-trivial phenomena in its non-equilibrium dynamics. Here, we demonstrate measurement-feedback control of a micromechanical resonator by driving the second-order nonlinearity (i.e., parametric squeezing) and directly measuring quadratic observables, which are given by the Schwinger representation of pseudo angular momentum (referred as Schwinger angular momentum). In contrast to that the parametric divergence occurs when the second-order nonlinearity is blindly driven, our measurement-feedback protocol enables us to avoid such a divergence and to achieve a strong noise reduction at the level of $-5.1\pm 0.2$ dB. This strong noise reduction originates in the effective cooling included in our measurement-feedback protocol, which is unveiled by investigating entropy production rates in a coarse-grained model. Our results open up the possibility of not only improving noise-limited sensitivity performance but also investigating entropy production in information thermodynamic machines with nonlinear measurement and feedback.
\end{abstract}

\section{Introduction}
Measurement-feedback control of fluctuation in mesoscopic systems has attracted large interest not only to precisely manipulate and stabilize the system but also to investigate stochastic dynamics itself under measurement and feedback. The presences of a measurement-feedback controller modifies the balance of entropy flows (i.e., the second law in the total system), and allows us to extract finite work to effectively heat up or cool down the system \cite{parrondo2015thermodynamics}. Such stochastic thermodynamics under the measurement and feedback have been investigated in various types of mesoscopic systems with artificial systems as well as natural biological ones \cite{ciliberto2017experiments}. In particular, micromechanical systems, such as an optically trapped nanoparticles \cite{andrieux2007entropy,toyabe2010experimental,gieseler2014dynamic} and micro/nanomechanical resonators \cite{douarche2005experimental,bonaldi2009nonequilibrium,serra2016mechanical,rossi2020experimental}, have been widely used to investigate the stochastic thermodynamics because fluctuation of their displacement (or velocity) can be precisely detected and controlled to implement measurement-feedback protocols.
 
So far, the linear measurement-feedback protocols, which consists of measurement and feedback of linear observables in phase space, have been demonstrated in micromechanical resonators to damp and amplify their displacement \cite{cohadon1999cooling, kleckner2006sub, poggio2007feedback}. Although such a linear measurement-feedback protocol can be simply implemented in a closed-loop setup, recently, non-trivial entropy production in that system with a finite delay has been unveiled in both theory \cite{munakata2014entropy,horowitz2014second} and experiment \cite{debiossac2020thermodynamics}. As a natural but further extension, bringing nonlinear observables to the measurement and feedback would extend these frameworks to be more general and non-trivial because nonlinearity naturally contains unique dynamical (e.g., instability) and stochastic (e.g., non-Markovianity) properties \cite{lifshitz2008nonlinear}. Although micro/nanomechanical resonators have been individually used to investigate both nonlinear control \cite{rugar1991mechanical,mahboob2016electromechanical,seitner2017parametric} and nonlinear measurement \cite{brawley2016nonlinear,leijssen2017nonlinear,asano2019optically}, combining them to develop a nonlinear measurement-feedback protocol has not been reported yet.

In this study, we propose and demonstrate continuous measurement-feedback control of a micromechanical resonator based on a quadratic observable, which is referred as ``Schwinger angular momentum'' because it is the quadratic form defined in the angular-momentum representation of bosons \cite{schwinger1965angular}. Because the Schwinger angular momentum holds quadratic and symmetric properties with SU(1,1) Lie algebra, we develop a continuous measurement-feedback protocol by combining parametric nonlinearity and measurement nonlinearity. The measurement nonlinearity enables us to directly readout the component of Schwinger angular momentum via nonlinear optomechanical transduction. The parametric nonlinearity enables us to drive the Schwinger angular momentum and squeeze the noise deviation (i.e., noise compression along a quadrature and noise amplification along an orthogonal one). In contrast to the blind parametric driving without measurement, our continuous measurement-feedback protocol enables us to achieve a non-equilibrium steady state (NESS) with strong noise squeezing at the level of $-5.1\pm 0.2$ dB. This is because a net cooling effect can avoid the divergence in parametric squeezing, which limits the level at -3 dB. This net cooling effect is unveiled by investigating the entropy production rates in our measurement-feedback protocols with a coarse-grained model.

In section 2, a theoretical framework for our quadratic measurement-feedback control is introduced with explicit expressions of entropy production rates in a thermal bath and a feedback controller in terms of Schwinger angular momentum in SU(1,1) Lie algebra. In section 3, the experimental setup and results are presented with analysis of the noise reduction level and entropy production rates by comparing our protocol with a random pumping protocol. In section 4, the possibility of controlling pure thermal motion with a higher order and intermodal regime with an optomechanical approach is discussed. Moreover, we discuss the importance of nonlinear measurement-feedback protocol in terms of stochastic thermodynamics.

\section{Theory}
\subsection{Parametric squeezing in Schwinger-angular-momentum space}
Parametric squeezing reduces noise deviation along a quadrature and amplifies it along the orthogonal quadrature in rotating-framed phase space. This parametric squeezing with continuous drive can be represented with an effective Hamiltonian $\mathcal{H}_\mathrm{eff}=G_0 (p^2-q^2)/4$, where $q$ and $p$ are the linear quadrature, and $G_0$ is strength of parametric driving (see Appendix A). Thus, the canonical equations for parametric squeezing are given by
\begin{equation}
\dot{q}=\frac{G_0}{2} p,\hspace{20pt}\dot{p}=\frac{G_0}{2} q.
\end{equation}

This quadratic form $(p^2-q^2)/4$ in the effective Hamiltonian is proportional to one of the Schwinger-angular-momentum components:
\begin{equation}
K_x=\frac{qp}{2},\hspace{10pt}K_y=\frac{q^2-p^2}{4},\hspace{10pt}K_z=\frac{q^2+p^2}{4}.
\end{equation}
Note that we fixed the phase in parametric drive so that the effective Hamiltonian contains only $K_y$ with $\theta_\mathrm{drive}=\pi/2$ from its general representation given by $\mathcal{H}_\mathrm{eff}=-G_0(K_x\cos\theta_\mathrm{drive}+K_y\sin\theta_\mathrm{drive})$. These three components satisfy $K_x^2+K_y^2-K_z^2=0$. This indicates that the dynamics of a mechanical resonator, which is commonly represented at a point in the phase space $(q,p)$, is shown as a dynamics on a hyperboloid in the Schwinger-angular-momentum space [see Fig 1(a)]. 

In addition to the geometrical property, they also satisfy a Lie algebra in SU(1,1) group such that
\begin{equation}
\left\{K_x,K_y\right\}=-K_z,\hspace{10pt}\left\{K_y,K_z\right\}=K_x,\hspace{10pt}\left\{K_z,K_x\right\}=K_y,
\end{equation}
where $\{A,B\}$ denotes the Poisson bracket defined by $\{A,B\}\equiv \partial A/\partial q \partial B/\partial p-\partial A/\partial p \partial B/\partial q$. Since the relation $\left\{K_x,K_y\right\}=-K_z$ only contains a negative sign compared with the other two, the effective Hamiltonian for parametric squeezing, $\mathcal{H}_\mathrm{eff}=-G_0 K_y$, leads a pseudo rotation around the $K_y$ axis in the Schwinger-angular-momentum space [see arrows in Fig. 1(a)]. This pseudo rotation is also confirmed in the dynamics of each component:
\begin{equation}
\dot{K}_x=G_0 K_z, \hspace{10pt}\dot{K}_z=G_0 K_x, \hspace{10pt}\dot{K}_y=0.
\end{equation}
Apparently, the general solutions of $K_x$ and $K_z$ are given by hyperbolic sine and cosine functions. 

The above formulation is valid to describe the stochastic dynamics of mechanical resonators with a probability density function $\mathcal{P}(q, p)$ in its phase space and $\mathcal{P}(K_x, K_y, K_z)$ in the Schwinger-angular-momentum space. In the phase-space description, the parametric squeezing with $\mathcal{H}_\mathrm{eff}$ amplifies the deviation along a diagonal quadrature, $q_+\equiv (q+p)/\sqrt{2}$, and reduces the deviation along its orthogonal portion, $q_- \equiv (q-p)/\sqrt{2}$ direction [see Fig. 1(b)]. On the other hand, in the Schwinger-angular-momentum description, the pseudo rotation with $\mathcal{H}_\mathrm{eff}$ around the $K_y$ axis leads to a biased probability distribution along the $K_z+K_x$ direction [see Fig. 1(c)]. Importantly, this pseudo rotation can be decomposed into noise compression along $K_z-K_x$ and noise expansion along $K_z+K_x$. By taking into account that
\begin{equation}
\langle K_z\pm K_z\rangle =\sigma (q_\pm)^2/2, \label{av_dev}
\end{equation}
the compression (expansion) along the $K_z-K_x$ ($K_z +K_x$) is regarded as the noise reduction (amplification) in the phase space. The divergence, which intrinsically limits the noise reduction level at -3 dB \cite{rugar1991mechanical}, appears as an infinite large noise amplification (i.e., noise expansion along $K_z+K_x$ direction in the pseudo rotation) when the drive strength $G_C$ is equivalent to the mechanical damping factor $\Gamma$. 
\begin{figure}[htb]
\includegraphics[width=15.0cm]{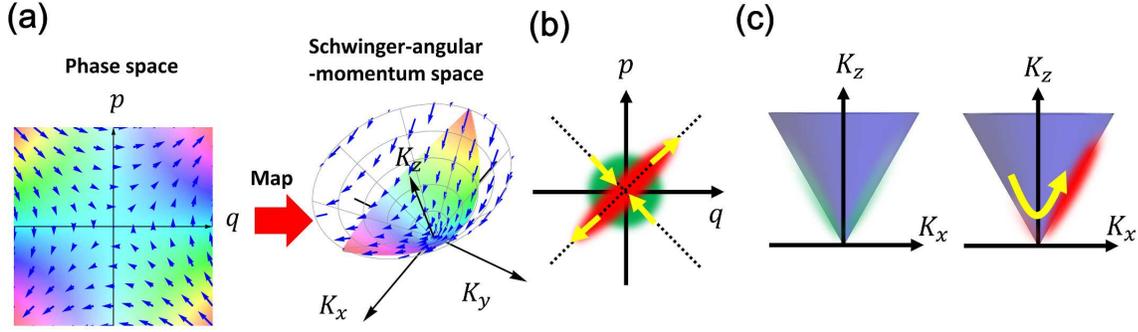}
\caption{\label{fig1}(a) Conceptual illustration of the phase space spanned by $q$ and $p$, and the Schwinger-angular-momentum space spanned by $K_x$, $K_y$, and $K_z$. The color map shows correspondence between the phase space and the Schwinger-angular-momentum space. The blue vectors show the force field of the parametric squeezing with $\mathcal{H}_\mathrm{eff}$. (b) and (c) Schematic of probability distribution of thermal equilibrium (green) and steady state with continuous squeezing (red) in  the phase space and the Schwinger-angular-momentum space, respectively. }
\end{figure}

\subsection{Protocol}
Because the divergent contribution in parametric drive can be distinguished with the sign of $K_x$, to avoid the divergence, our measurement-feedback protocol is derived as a switching operation of the parametric drive with respect to this sign. Switching on the parametric drive (i.e., $G_0$ takes a non-zero value) only when $K_x <0$ leads a suppression of the probability density function in the Schwinger-angular-momentum space. On the other hand, switching it off ($G_0=0$) when $K_x >0$ can avoid the divergence in parametric driving. 

To read out the sign of $K_x$ to construct a feedback loop, we can utilize nonlinear optomechanical transduction in which higher harmonic signals are generated thanks to a dispersive modulation of optical phase via mechanical motion \cite{brawley2016nonlinear,leijssen2017nonlinear,asano2019optically}. In particular, the sine and cosine parts in the second-order harmonics are regarded as $K_x$ and $K_y$, respectively. To achieve steady-state squeezing, our measurement-feedback protocol is continuously repeated with $K_x$ directly measured via nonlinear optomechanical transduction and the parametric drive switched on or off with respect to the sign of measured $K_x$ [see Fig. 2(a)].
\begin{figure}[htb]
\includegraphics[width=15.0cm]{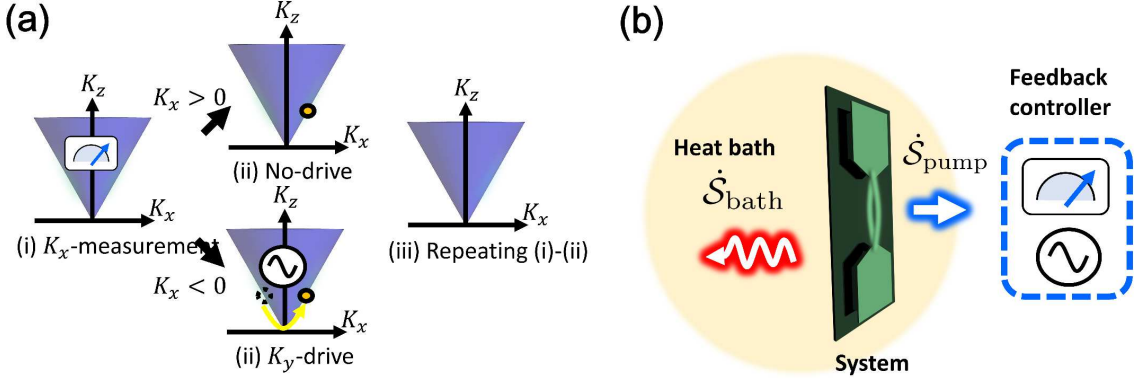}
\caption{\label{fig2} (a) Schematic of the measurement-feedback protocol in the Schwinger-angular-momentum space. Initially, the sign of $K_x$ is measured via nonlinear optomechanical transduction (i). Then, only if $K_x<0$, the parametric drive is switched on (ii). By repeating these two processes, we achieve a non-equilibrium steady state (NESS) that is partially distributed around only $K_x>0$ (iii). (b) Conceptual illustration of net cooling via measurement-feedback protocol. In addition to the entropy production in the thermal bath with its rate $\dot{\mathcal{S}}_\mathrm{bath}$, there exists entropy production called ``entropy pumping'' with the rate $\dot{\mathcal{S}}_\mathrm{pump}$ due to the existence of the feedback controller. The vectors describe the actual directions of both fluxes when the measurement-feedback protocol succeeds, where $\dot{\mathcal{S}}_\mathrm{bath}\geq 0$ and $\dot{\mathcal{S}}_\mathrm{pump}\geq 0$ in our definition.}
\end{figure}

\subsection{Entropy production rates in a feedback controller}
The parametric squeezing is intrinsically a heating operation because noise amplification levels along a quadrature are always larger than noise reduction levels along the orthogonal ones, which can be readily confirmed by taking into account the Shannon entropy in NESS (Appendix B). Thus, avoiding the heating part of pseudo rotation [i.e., $K_x>0$ in Fig. 1(c)] in Schwinger-angular-momentum space with our measurement-feedback protocol effectively induces a net cooling effect. To unveil this net cooling effect hidden in our protocol, we take into account entropy production in our setup, which consists of a system (a mechanical resonator), a heat bath, and a feedback controller [see Fig.2(b)]. Stochastic thermodynamics provides us insights into heating and cooling in a non-equilibrium steady state in terms of the balance among entropy production rates, where additional entropy production exists due to the presence of a feedback controller \cite{sagawa2012nonequilibrium,munakata2012entropy,horowitz2014second}. In particular, entropy production with a continuous measurement-feedback controller has been investigated by coarse-graining methods, where the degree of feedback memory is coarse-grained \cite{munakata2012entropy,munakata2014entropy,horowitz2014second}. Such investigations have indicated the existence of an additional entropy production called ``entropy pumping" due to the existence of a feedback controller. Because previous studies on the entropy pumping have focused on a linear operation (i.e., observable and controlled quantities directly correspond to a quadrature in its phase space) \cite{munakata2012entropy,munakata2014entropy,horowitz2014second}, we attempt to formulate the entropy production rate in our quadratic measurement-feedback protocol with respect to the Schwinger angular momentum as follows.

To achieve the expression of entropy productions with the rotating wave approximation, we start from the stochastic dynamics of the mechanical displacement $X$ and momentum $P$ in the laboratory frame with a measurement-feedback operation given by the Langevin equations
\begin{eqnarray}
&\dot{X}=P,\\
&\dot{P}+\Gamma P+\Omega^2 X-2G_0\Omega f(M)\cos 2\Omega t X=F_\mathrm{th},\label{fullLang}
\end{eqnarray}
where $\Gamma$ is the damping rate, and $\Omega$ is the angular frequency of the mechanical resonator. Note that we set the effective mass to be unity in the following discussion for simplicity (replacing $k_B $ by $k_B/m_\mathrm{eff}$ provides us the exact expressions with the effective mass of mechanical resonator, $m_\mathrm{eff}$). The Langevin force $F_\mathrm{th}$ satisfies $\langle F_\mathrm{th}(t')F_\mathrm{th}(t)\rangle=2k_BT\Gamma \delta(t-t')$ with the Boltzmann constant $k_B$ and the temperature of the environment $T$. The protocol of measurement and feedback is expressed by the feedback function $f(M)$ with the memory value $M$. In our protocol, the feedback function is given by a Heaviside function $\theta(\cdot)$ as $f(M)=\theta(M)$. Analytical difficulty in the discontinuity of the Heaviside function is avoided by taking into account the finite measurement noise, which is assumed as Gaussian white noise. Thus, the conditional probability without feedback delay,
\begin{equation}
\mathcal{P}(M|\mathcal{M}(X,P))=\frac{1}{\sqrt{2\pi\sigma_M^2}}\exp\left[-\frac{(\mathcal{M}(X,P)-M)^2}{2\sigma_M^2}\right], \label{Pcond}
\end{equation}
is suitable for modeling our measurement-feedback loop, where $\mathcal{M}(X,P)$ is the target observable in the measurement, and $\sigma_M$ is the standard deviation of the Gaussian noise. By multiplying Eq. (\ref{Pcond}) by Eq. (\ref{fullLang}) and integrating both sides with respect to $M$, a coarse-grained dynamics is represented as follows:
\begin{eqnarray}
\dot{P}+\Gamma P+\Omega^2 X-G_0\Omega \mathrm{erfc}\left(\frac{\mathcal{M}(X,P)}{\sqrt{2}\sigma_M}\right)\cos 2\Omega t X=F_\mathrm{th},\label{CoarseLang}
\end{eqnarray}
where $\mathrm{erfc}(\cdot)$ is a complementary error function reflecting the switching operation with finite measurement noise. The entropy production in the continuous measurement and feedback is formulated from this coarse-grained Langevin equation. Here, note that we formulate the entropy production rates by utilizing both a path integral formalism \cite{imparato2006fluctuation,munakata2012entropy,munakata2014entropy}, which allows us to formulate them with clear physical meaning in our model (Appendix C), and probability currents \cite{tome2010entropy,spinney2012entropy,horowitz2014second}, which might be better for confirming the former results (Appendix D). The two formalisms result in the same expression of the entropy production rates with respect to the Schwinger angular momentum. The formulated entropy production rate is decomposed to two contributions: the entropy production in the thermal bath, $\dot{\mathcal{S}}_\mathrm{bath}$, due to the heat flux from the system, and the entropy production thanks to the existence of the feedback controller, $\dot{\mathcal{S}}_\mathrm{pump}$, i.e. ``entropy pumping" owing to the measurement and feedback as follows:
\begin{eqnarray}
\left\langle \dot{\mathcal{S}}_\mathrm{bath}\right\rangle&\approx \frac{2\Omega^2\Gamma}{k_BT}\left(\langle K_z \rangle-\frac{k_BT}{2\Omega^2}\right),\label{Sbath_main}\\
\left\langle \dot{\mathcal{S}}_\mathrm{pump}\right\rangle&\approx \sqrt{\frac{1}{2\pi}} \frac{\tilde{G}_0\Gamma}{\sigma_M}\left\langle K_z\exp\left(-\frac{K_x^2}{2\sigma_M^2}\right)\right\rangle,\label{Spump_main}
\end{eqnarray}
where $\langle\cdot\rangle$ denotes the stochastic average, and $\tilde{G}_0=G_0/\Gamma$ is the effective driving strength as a dimensionless quantity. Note that the entropy production rates in the thermal bath, $\dot{\mathcal{S}}_\mathrm{bath}$, is defined to be positive when the heat flux flows out of the system; and $\dot{\mathcal{S}}_\mathrm{pump}$ is defined to be positive when the feedback controller pulls the entropy from the system [please see the vectors in Fig. 2(b)].

The entropy production in the thermal bath is given by the shift of the vertical component of Schwinger angular momentum $K_z$ from its value in the equilibrium because $K_z$ directly corresponds to the oscillation energy (phonon number) in the resonator. On the other hand, the entropy pumping rate is given by a function of $K_x$, $K_z$, and the measurement noise characterized by its deviation of $\sigma_M$. Because of $K_z\geq 0$, $\langle \dot{\mathcal{S}}_\mathrm{pump}\rangle\geq 0$ always holds, which indicates that the entropy pumping operates to pull the entropy from the system. These three contributions satisfy the second-law-like inequality
\begin{equation}
\langle \dot{\mathcal{S}}_\mathrm{bath}\rangle+\langle\dot{\mathcal{S}}_\mathrm{pump}\rangle \geq 0 \label{ineq3}.
\end{equation}
This inequality implies that $\dot{\mathcal{S}}_\mathrm{bath}$ may take a negative value in contrast to the case of no-feedback operation ($\langle \dot{\mathcal{S}}_\mathrm{bath}\rangle\geq 0$). In this negative regime, the system can operate as a cooler where heat is pumped out of the thermal bath to the feedback controller. 

\section{Experiment}
\subsection{Setup}

Our measurement-feedback protocol was implemented on a measurement-feedback loop with a micromechanical resonator [see Fig. 3(a)]. A doubly-clamped silicon nitride mechanical resonator (150 $\mathrm{\mu m}$-long, 5-$\mathrm{\mu}$m-wide, and 525-nm-thick) was fabricated via thermal chemical vapor deposition, and placed in a vacuum environment ($\sim 10^{-4}$ Pa). The resonator showed a high quality factor of $3.0\times 10^4$ in its fundamental flexural mode at the frequency of $\Omega=2\pi \times 510$ kHz at room temperature. Linear quadratures ($q$ and $p$) and a component of Schwinger angular momentum ($K_x^M$) in its mechanical motion were extracted from a laser Doppler interferometer (LDI). Here, we denote the directly measured Schwinger-angular-momentum component with $K_x^M$ to distinguish it from the one, $K_x^P=qp/2$, calculated via post-processing with the measured $q$ and $p$. The output of LDI was connected to lock-in amplifiers with a reference frequency of $\Omega$ for the linear quadratures and that of $2\Omega$ for the Schwinger-angular-momentum components. Note that we induced additional white noise via an piezoelectric sheet attached on the resonator substrate to improve the signal-to-noise ratio for $K_x^M$ (the effective temperature was estimated to be $T_\mathrm{eff}\approx 10^5$ K). By once connecting the output from LDI to a spectrum analyzer, spectrum at $2\Omega$, which reflects the signal of $K_x^M$, was observed as well as that at $\Omega$  [see Fig. 3(b)]. The linewidth at $2\Omega$ becomes twice of that at $\Omega$ because $K_x^M$ is the quadratic observable in $q$ and $p$. The measured component, $K_x^M$, was used to switch on (off) the parametric pump with an oscillation frequency of $2\Omega$ when the component $K^M_x$ is negative (positive) with a radio-frequency switch. To pump the mechanical resonator along the $K_y$-direction, the phase of the parametric pump was 90-degrees shifted from that for the reference signal to the lock-in amplifier for measuring $K_x^M$. The parametric pump was fedback to the mechanical resonator via the piezoelectric sheet. A typical temporal response in this measurement-feedback loop is shown in Fig. 3(c) where the parametric pump (red curve) with $2\Omega$ was turned on when $K^M_x$ (green plot) was negative. Note that the time constant of the lock-in amplifier was fixed at $\tau_L=100$ $\mathrm{\mu sec}$ to achieve all information on mechanical motion with the mechanical lifetime $\tau_M=59$ msec.

\begin{figure}[htb]
\includegraphics[width=15.0cm]{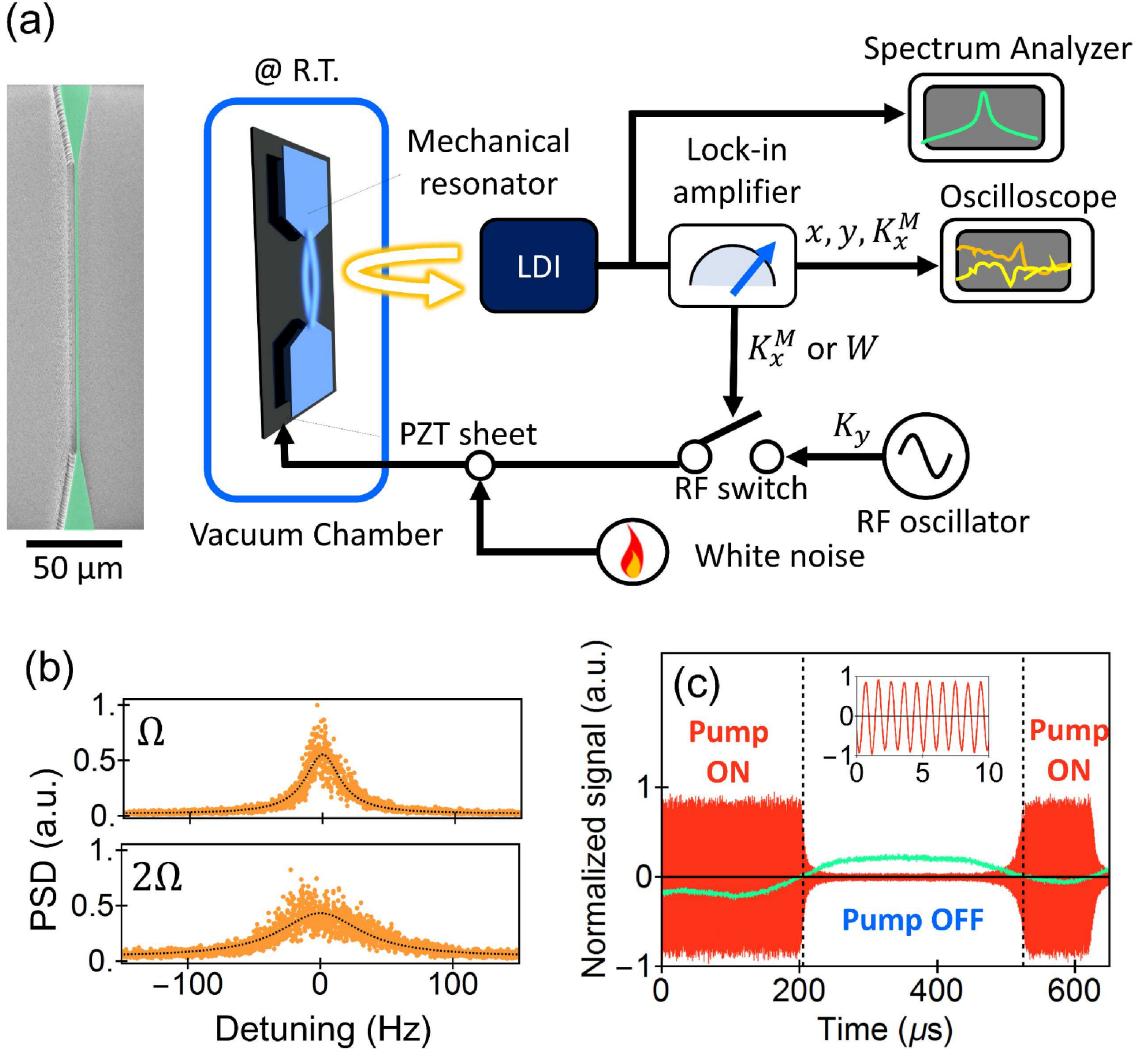}
\caption{\label{fig3} (a) Schematic of experimental setup with a high-Q silicon-nitride doubly-clamped beam. A radio-frequency (rf) oscillator with twice of the mechanical frequency was connected to an rf switch for the parametric driving. This switching operation was determined with respect to the measurement outcome of $K^M_x$ from a laser Doppler interferometer (LDI). The mechanical excitation was done by the parametric drive signal from the rf oscillator and white noise signal via a piezoelectric signals. Here, note that this interferometer also yields temporal data of the phase quadratures $q$ and $p$ recorded on the oscilloscope. (b) Frequency spectra at $\Omega$ (measurement of phase quadratures) and $2\Omega$ (measurement of Schwinger angular momenta). (c) Typical temporal sequence of our measurement-feedback protocol. The parametric drive signal with $2\Omega$ frequency (red curve) was sent to the piezoelectric sheet only if the measurement outcome (the signal of $K^M_x$) took negative values. The inset shows the enlarged data which explicitly indicate the period of $1/(2\Omega)$.}
\end{figure}

\subsection{Noise reduction level and NESS in Schwinger-angular-momentum space}
An initial equilibrium probability distribution was observed without any feedback drive as an equally-distributed Gaussian distribution. To verify how our protocol affects the probability distribution, we demonstrated both random pumping (i.e., the parametric drive was randomly switched on or off) and our measurement-feedback protocol, and evaluated noise reduction and amplification levels in their non-equilibrium steady states. Here, the noise reduction level, $\xi_\mathrm{red}$, is defined as $\mathrm{min}_\theta \sigma(q_{\theta,\mathrm{NESS}})/\sigma(q_{\mathrm{INIT}})$, where $\sigma(\cdot)$ is the standard deviation, $q_\mathrm{INIT}$ is the quadrature in the initial equilibrium, and $q_\theta=q\cos\theta+p\sin\theta$ is the quadrature with an arbitrary angle $\theta$. The noise amplification level, $\xi_\mathrm{amp}$, is defined as the deviation along the orthogonal part, i.e., $\mathrm{min}_\theta\sigma(\bar{q}_{\theta,\mathrm{NESS}})/\sigma(q_{\mathrm{INIT}})$, where $\bar{q}_\theta=-q\sin\theta+p\cos\theta$. In the case of the random pumping [see Fig. 4(a)], a squeezed Gaussian distribution with $\xi_\mathrm{red}\leq \xi_\mathrm{amp}$ was achieved in the same way as in standard noise squeezing. However, the noise reduction level was limited to about -3 dB around the drive voltage of 150 mV [see Fig. 4 (b)]. Drive voltages larger than 150 mV induced parametric instability where both  $\xi_\mathrm{red}$ and $\xi_\mathrm{amp}$ increased. On the other hand, once our measurement-feedback protocol was demonstrated, a non-equilibrium steady state with $\xi_\mathrm{red}\geq \xi_\mathrm{amp}$ was observed with a non-Gaussian probability distribution [see Fig. 4(c)]. This non-Gaussianity directly reflects the non-Gaussian properties of our quadratic observables \cite{nha2006entanglement}. Moreover, the noise reduction level finally reached $-5.1\pm 0.2$ dB over the -3 dB, whereas the noise amplification level was suppressed at $1.7\pm 0.2$ dB [see Fig. 4(d)].  

\begin{figure}[htb]
\includegraphics[width=15.0cm]{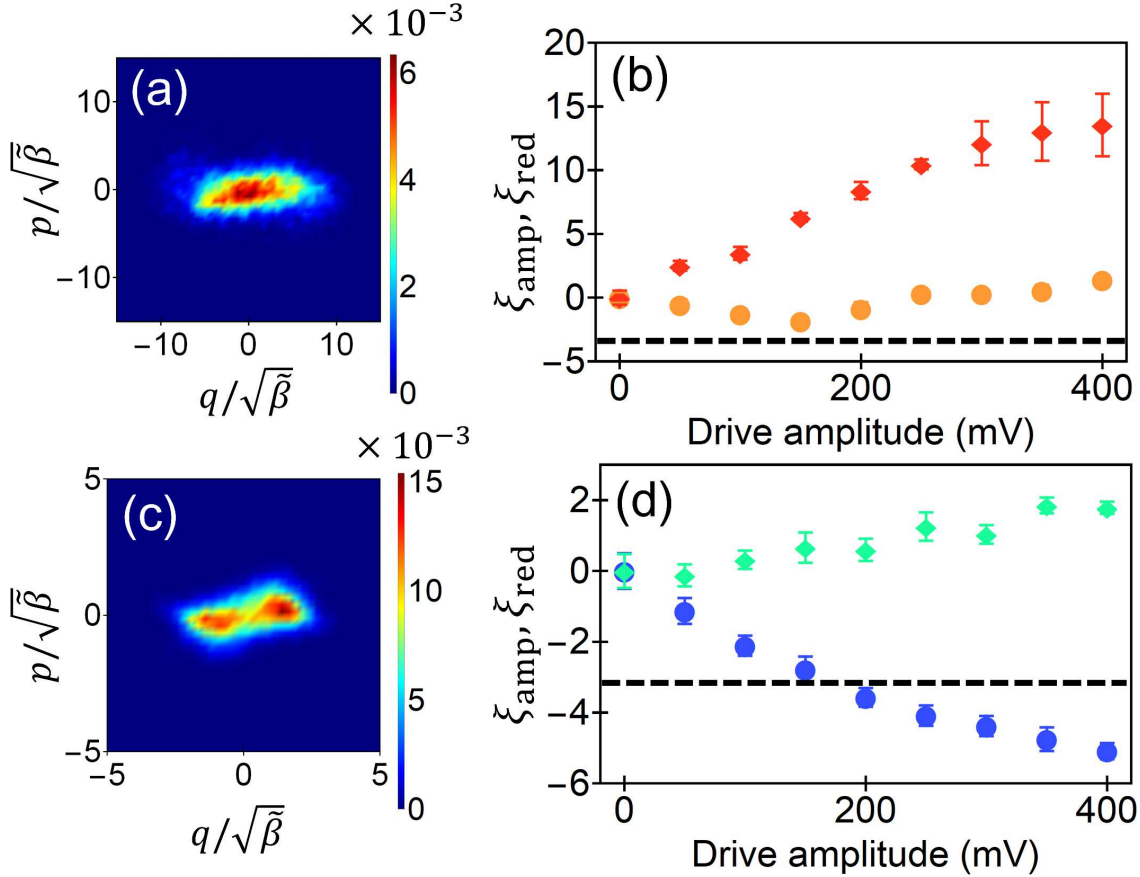}
\caption{\label{fig4} (a) and (c) Probability density functions in the phase space with the random protocol and our measurement-feedback  protocol, respectively. (b) and (d) The maximum and minimum standard deviations of these distributions are evaluated with respect to the parametric drive amplitudes The vertical error-bar corresponds to the standard deviation in ten trials.}
\end{figure}

Because we attempt to perform fully quadratic measurement-feedback protocol on the Schwinger-angular-momentum space, the NESS in that space is totally different between our measurement-feedback protocol and the random protocol. In contrast to $K_x^M$, we use the dataset of $q$ and $p$ to evaluate Schwinger angular momentum $K_i^P$ ($i=x,y,z$) via the post-processing to reconstruct the probability density function in the Schwinger-angular-momentum space. Note that $K_i^P$ has a more accurate value than $K_i^M$ because the signal-to-noise ratio in $\Omega$ is better than that in $2\Omega$, although $K_i^P$ was only available in the post-processing without any fast feedback processor. Thus, we used $K_x^M$ in the measurement-feedback control and $K_x^P$ in the analysis to, for instance, calculate the stochastic average of $K_x$. Figure 5 shows marginal probability density functions in the space spanned by $K_x^P$ and $K_z^P$ for both the random protocol and our measurement-feedback protocol. Apparently, the initial equilibrium state was isotropically distributed [see Fig.5 (a) and (d)]. On the other hand, the non-equilibrium steady states in the random protocol [see Fig. 5(b) and (c)] and our measurement-feedback protocol [see Fig. 5(e) and (f)] show biased distributions along $K_x>0$ directions due to the pseudo rotation in parametric squeezing. We emphasize that the axes scales are totally different between the random protocol and our measurement-feedback protocol. Although the distribution is broadened in the random protocol because the parametric drive is blindly injected, the distribution in our protocol in the positive $K_x$ side is almost completely kept thanks to the switch-off operation. 
\begin{figure}[htb]
\includegraphics[width=15.0cm]{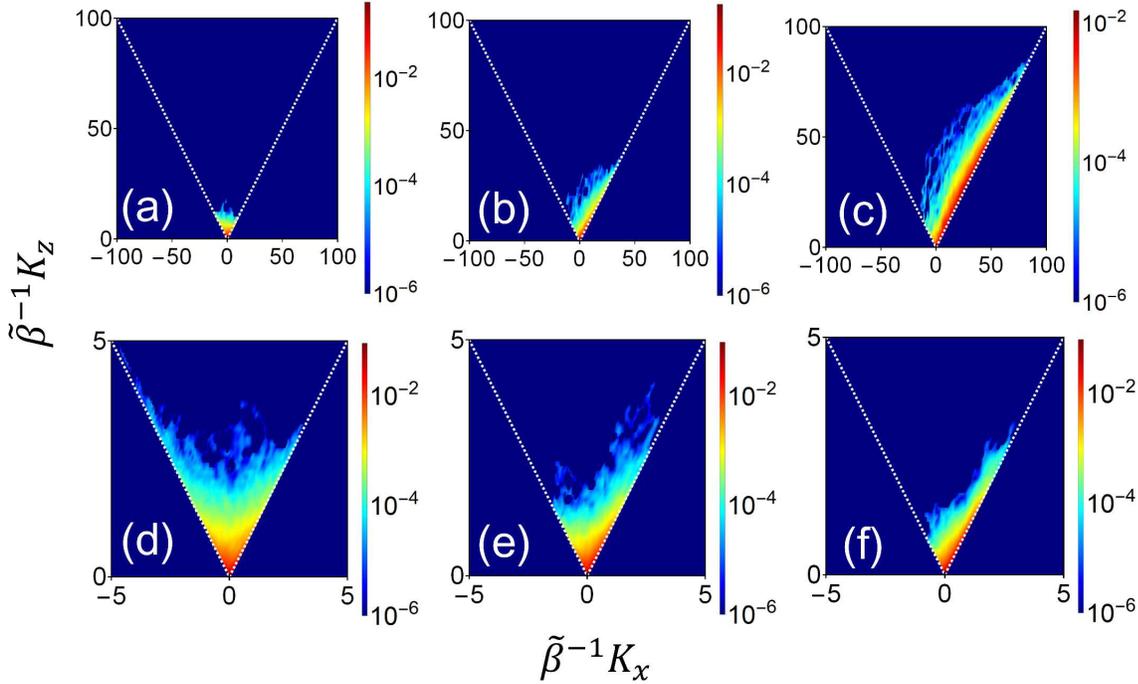}
\caption{\label{fig5} Probability density functions with the random protocol with the drive amplitudes of (a) 0, (b) 150, and (c) 400 mV, and with the feedback protocol in the drive amplitude of (d) 0, (e) 150, and (f) 400 mV. Note that the axes scales are totally different between the two protocols.}
\end{figure}

\subsection{Net cooling effect and entropy production rates}
To unveil the net cooling effect hidden in our measurement-feedback experiment, we evaluate the entropy production rate from the heat bath, $\dot{\mathcal{S}}_\mathrm{bath}$, and the entropy pumping rate $\dot{\mathcal{S}}_\mathrm{pump}$ from the theoretical formulation in Eqs. (\ref{Sbath_main}) and (\ref{Spump_main}). Note that the stochastic averages in Eqs. (\ref{Sbath_main}) and (\ref{Spump_main}) are determined by using $K_i^P$. $\dot{\mathcal{S}}_\mathrm{bath}$ is experimentally determined by
\begin{equation}
\langle\dot{\mathcal{S}}_\mathrm{bath}\rangle=\Gamma \frac{\langle K^P_z\rangle-\langle K_z^P\rangle_0}{\langle K_z^P\rangle_0},
\end{equation}
where $\langle \cdot\rangle_0$ denotes the stochastic average in the initial equilibrium state. In the same manner, we can obtain
\begin{eqnarray}
\langle\dot{\mathcal{S}}_\mathrm{pump}\rangle=\sqrt{\frac{1}{2\pi}}\frac{\Gamma \tilde{G}_0}{\sigma_M}\left\langle K^P_z\exp\left[-\frac{(K^P_x)^2}{2\sigma_M^2}\right]\right\rangle.
\end{eqnarray}
To estimate the entropy production rates, the effective drive strength $\tilde{G}_0$ and measurement noise deviation $\sigma_M$ are required. The $\tilde{G}_0$ was estimated from the noise reduction level in the random protocol shown in Fig. 4(b) (yellow circles). From the divergence condition $\tilde{G}_0=2$ in the random protocol, at which the noise reduction level changes from decreasing to increasing, we can determine $\tilde{G}_0/V_\mathrm{drive}=10.4$ $\mathrm{V}^{-1}$. The $\sigma_M$ was estimated to be $0.52\pm 0.07$ from both $K_x^M$ and $K_x^P$ by assuming the Gaussian noise in measurement (see Appendix E). 

The entropy production rates normalized by the mechanical damping rate $\Gamma$ are shown in Fig. 6 (a). With increasing driving strength $\tilde{G}_0$, the entropy pumping rate, $\dot{\mathcal{S}}_\mathrm{pump}$, naturally increases with positive values. On the other hand, the entropy production rate in the thermal bath, $\dot{\mathcal{S}}_\mathrm{bath}$, decreases with negative values. This indicates that our measurement-feedback protocol successfully induces a net cooling effects where the system pulls the heat from the thermal bath to the feedback controller [see Fig.6 (b)]. Thus, this net cooling effect avoids heating due to the parametric driving, and results in stronger squeezing in its NESS because it allows us to inject stronger parametric driving over the limitation due to the divergence. Moreover, we can confirm the second-law-like inequality $\langle \dot{\mathcal{S}}_\mathrm{bath}\rangle+\langle \dot{\mathcal{S}}_\mathrm{pump}\rangle\geq0$ from the experimental results. The total entropy production rate, $\langle \dot{\Sigma}\rangle \equiv \langle \dot{\mathcal{S}}_\mathrm{bath}\rangle+\langle \dot{\mathcal{S}}_\mathrm{pump}\rangle$, apparently increases with increasing $\tilde{G}_0$. This is because the smaller $\tilde{G}_0$ purely induces the parametric squeezing with the net cooling [see Fig. 6(b)], while the larger $\tilde{G}_0$ intrinsically induces an additional heating in its pseudo rotation despite successful operation of the protocol [see Fig. 6(c)].

\begin{figure}[htb]
\begin{center}
\includegraphics[width=10cm]{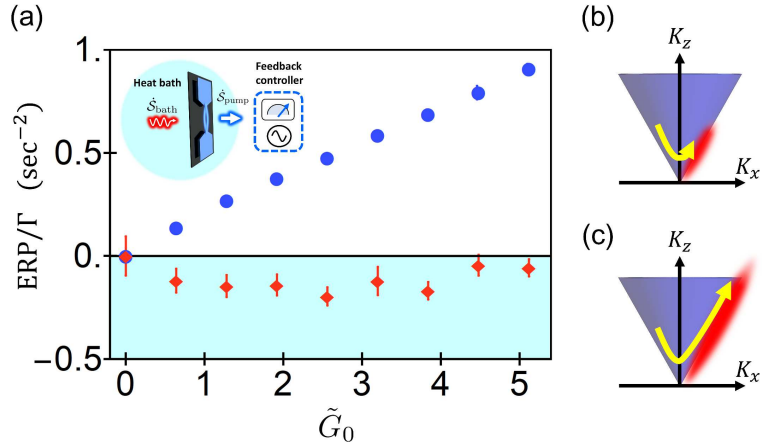}
\caption{\label{fig6} (a) Entropy production rate (EPR) normalized by the mechanical damping rate $\Gamma$ in our measurement-feedback protocol. The red diamonds show the entropy production rates in the thermal bath, $\langle\dot{\mathcal{S}}_\mathrm{bath}\rangle/\Gamma$, and the blue dots show that in the feedback controller, $\langle\dot{\mathcal{S}}_\mathrm{pump}\rangle/\Gamma$, (i.e., entropy pumping rate). The blue shaded area corresponds to the cooling regime where the system operates as a cooler (conceptual image is shown in the inset). The error bars show the standard deviation in ten trials. (b) and (c) Schematic of pseudo rotation with (b) smaller $G_0$ and (c) larger $G_0$ in the Schwinger angular momentum space.}
\end{center}
\end{figure}

It is intuitive that entropy production rates between our measurement-feedback protocol and the random protocol can be continuously related with respect to the measurement noise deviation $\sigma_M$ (i.e., $\sigma_M\to \infty$ corresponds to the random protocol). Thus, we numerically evaluate the entropy production rates $\langle\dot{\mathcal{S}}_\mathrm{bath} \rangle$ and $\langle\dot{\mathcal{S}}_\mathrm{pump} \rangle$ with different $\sigma_M=\{10^{-1},10^0,10^1\}$ (see Fig. 7), and compare them with the analytical expression of $\langle\dot{\mathcal{S}}_\mathrm{bath} \rangle$ in continuous and random driving (see Appendix F). As the measurement noise deviation increases, the slope of increment of $\langle\dot{\mathcal{S}}_\mathrm{bath} \rangle$ become steep while the slope of increment of $\langle\dot{\mathcal{S}}_\mathrm{pump} \rangle$ becomes gentle. This is because the measurement error induces the heating from parametric driving with the pseudo rotation in $K_x\geq 0$, which decreases the influence of measurement and feedback (it becomes close to the random protocol). We emphasize again that our measurement-feedback scheme yields a noise-squeezed NESS in the cooling regime, whereas the conventional parametric scheme (continuous or random) yields it in the heating regime. 
\begin{figure}[htb]
\includegraphics[width=15.0cm]{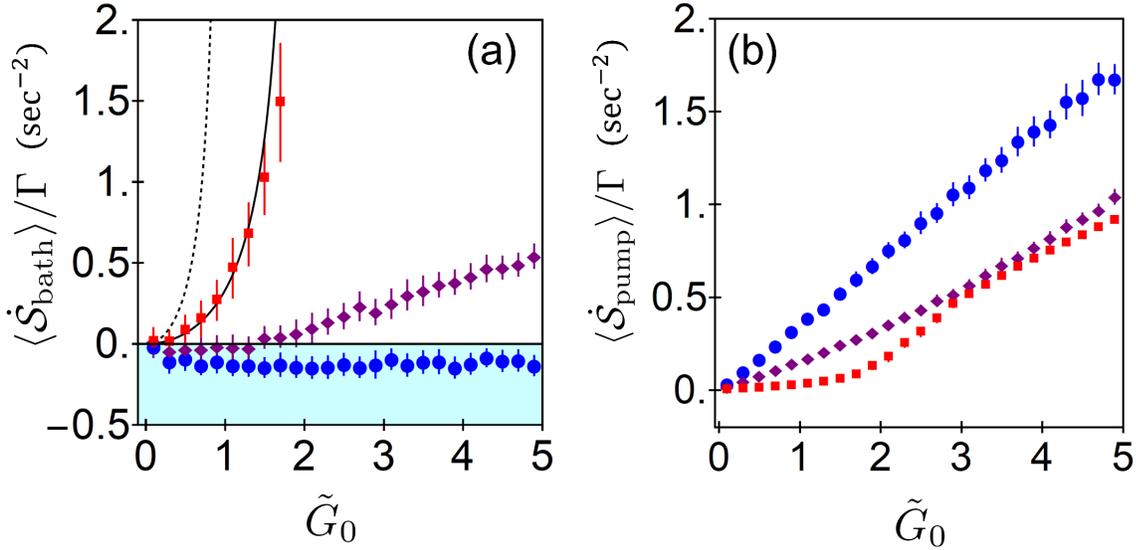}
\caption{\label{fig7} Numerically evaluated entropy production rates of (a) $\langle\dot{\mathcal{S}}_\mathrm{bath} \rangle/\Gamma$ and (b) $\langle\dot{\mathcal{S}}_\mathrm{pump} \rangle/\Gamma$. The dots, diamonds, and squares correspond to the noise deviation of $\sigma_M=$, $10^{-1}$, $10^0$, and $10^1$, respectively. The error bars show the standard deviation within fifty trials. The blue shaded area in (a) corresponds to the cooling regime.}
\end{figure}

\section{Discussion}

Our proof-of-principle experiment for the continuous measurement feedback control with Schwinger angular momentum was performed with additional white noise to improve the signal-to-noise ratio in the quadratic measurement in the Doppler interferometer. The quadratic measurement can be extended to pure thermal fluctuation in mechanical resonators with Doppler interferometry by increasing the mechanical Q factor and decreasing effective mass in the mechanical modes \cite{asano2019optically}. Although the mechanical Q factor simply contributes to the signal-to-noise ratio as $\sqrt{Q}$ dependence, the inverse of effective mass linearly contributes to it. Thus, mechanical resonators with small effective mass (e.g. graphene drum resonators \cite{bunch2007electromechanical,will2017high}) are suitable for performing our protocol with pure thermal motion with the Doppler interferometry. As an alternative approach, cavity optomechanical coupling in the unresolved sideband regime is also available for measuring higher order harmonics in mechanical modes, and has been demonstrated for observing them in pure thermal motion \cite{brawley2016nonlinear, leijssen2017nonlinear}. Furthermore, extension to more higher order observables and intermodal observables would open the way to more functionally control in the nonlinear measurement-feedback frameworks.

It is important to emphasize that experimental verification of the second-law-like inequality with the entropy pumping rate has been demonstrated only in linear measurement-feedback schemes \cite{debiossac2020thermodynamics}. In contrast, we investigated entropy production under fully nonlinear measurement-feedback control of stochastic dynamics with a certain symmetry (i.e., certain geometry of variable space). Utilizing such rich intrinsic and external (measurement) nonlinearity in mechanical resonators might further promote the experimental verification of various types of thermodynamic limitations with information resources \cite{shiraishi2018speed,ito2020stochastic}.

\section{Conclusion}
In conclusion, we have demonstrated measurement-feedback control of Schwinger angular momentum using a high-Q silicon nitride mechanical resonator by Doppler interferometry. A strong noise reduction level $-5.1\pm 0.2$ dB was achieved by suppressing the heating effect in parametric drive by well manipulating the probability distribution in Schwinger-angular-momentum space. Furthermore, the entropy production rate hidden in the quadratic measurement-feedback control was unveiled in both theory and experiment via the second-low-like inequality in the coarse-grained dynamics. This nonlinear measurement-feedback control framework for stochastic dynamics can be extended to the higher order and multimodal regime, which will be applicable to further experimentally investigate information thermodynamics with symmetry and nonlinearity.

\section*{Acknowledgement}
We thank Kensaku Chida for fruitful discussions. This work was partly supported by a MEXT Grant-in-Aid for Scientific Research on Innovative Areas (Grants No. JP15H05869).

\clearpage

\appendix

\section{Dynamics in the rotating frame with parametric squeezing}
The dynamics of displacement $X$ in a mechanical resonator with parametric force $F_\mathrm{p}$ is given by
\begin{equation}
\ddot{X}+\Gamma\dot{X}+(\Omega^2+F_\mathrm{p})X=F_\mathrm{th}
\end{equation}
where $\Gamma$ is the damping factor, $\Omega$ is the mechanical angular frequency, and the Langevin force $F_\mathrm{th}$ is given by $\langle F_\mathrm{th}(t) F_\mathrm{th}(t')\rangle=2k_BT\Gamma \delta(t-t')$. To induce the parametric squeezing, the parametric force has a double period of mechanical oscillation, i.e., $F_\mathrm{p}=-2G_0\Omega\cos2\Omega t$ with the drive strength of $G_0$. The dynamics in rotating phase space spanned by linear quadratures $q$ and $p$ ($X=q\cos\Omega t+p\sin\Omega t$) is approximated by $\ddot{z}\ll \Omega \dot{z}$ $(z=q, p)$ and $\Gamma/\Omega \ll 1$, that is the linear quadrature in a high-Q mode more slowly varies than the mechanical frequency as follows:
\begin{eqnarray}
&-\Omega(2\dot{q}+\Gamma q)\sin\Omega t+\Omega(2\dot{p}+\Gamma p)\cos\Omega t,\nonumber\\
&+G_0\Omega\left[q(\cos\Omega t+\cos 3\Omega t)+p(\sin 3 \Omega t-\sin\Omega t)\right]=F_\mathrm{th}.
\end{eqnarray}
To take into account the rotating term with $\Omega$, the Langevin force is split as $F_\mathrm{th}=-f_q\sin\Omega t-f_p\cos\Omega t$ where $\langle f_z(t)f_z(t')\rangle=2k_BT\Gamma\delta(t-t')$ is satisfied. This leads to the Langevin equations for each quadrature with the parametric squeezing as follows:
\begin{eqnarray}
\dot{q}=-\frac{\Gamma}{2} q+\frac{G_0}{2}p+\sqrt{\tilde{\beta}\Gamma}\xi_q,\\
\dot{p}=-\frac{\Gamma}{2} p+\frac{G_0}{2}q+\sqrt{\tilde{\beta}\Gamma}\xi_p,
\end{eqnarray}
where $\tilde{\beta}\equiv k_BT/\Omega^2$. Note that the effective rotating-framed Hamiltonian is given by $\mathcal{H}_\mathrm{eff}=G_0(p^2-q^2)/4$.

\section{Heating effect in parametric squeezing}
Heating in continuous parametric squeezing can be simply seen in the change in the Shannon entropy between the initial and final equilibrium states, which is given by
\begin{equation}
\Delta \mathcal{H}_\mathrm{S}=\frac{1}{2}\mathrm{ln}\frac{\left|\Sigma_f\right|}{\left|\Sigma_i\right|},
\end{equation}
where $|\Sigma_i|$ and $|\Sigma_f|$ are determinants of covariant matrices in the initial equilibrium state and final squeezed state, respectively. From Langevin equations for continuous parametric squeezing given by
\begin{eqnarray}
\dot{q}&=-\frac{\Gamma}{2} q+\frac{G_C}{2} p+\sqrt{\Gamma \tilde{\beta}}\xi_q,\\
\dot{p}&=-\frac{\Gamma}{2} p+\frac{G_C}{2} q+\sqrt{\Gamma \tilde{\beta}}\xi_p,
\end{eqnarray}
where $G_C$ is the strength of continuous parametric drive, the determinant of the covariant matrix in the final squeezed state is given by
\begin{equation}
\left|\Sigma_f\right|=\frac{16\tilde{\beta}\Gamma}{(\Gamma^2-G_C^2)^2}.
\end{equation}
This leads to
\begin{equation}
\Delta \mathcal{H}_\mathrm{S}=\mathrm{ln}\frac{1}{1-G_C^2/\Gamma^2}.
\end{equation}
Since the Shannon entropy monotonically increases in the stable squeezing regime $G_0<\Gamma$, the system (i.e., mechanical resonator) is totally heated up due to the parametric squeezing.

\section{Entropy production in our protocol by means of path integral}
Total entropy production $\Sigma$, which is always non-negative, is defined by the Kullback-Leibler divergence between the forward probability distribution and the inverse probability distribution as
\begin{eqnarray}
\Sigma=\ln \frac{\mathcal{P}_\mathrm{fwd}}{\mathcal{P}_\mathrm{inv}}\geq 0.
\end{eqnarray}
In the case of the continuous measurement-feedback control, entropy production has been investigated in a coarse-grained dynamics, where the memory degree of freedom in measurement is coarse-grained in its equation of motion \cite{munakata2012entropy,munakata2014entropy,horowitz2014second}. The inverse probability in the coarse-grained dynamics was defined as a probability with  ``conjugate" dynamics, in which the time-reversal parity of feedback cooling forces is defined to be positive \cite{munakata2012entropy,munakata2014entropy}. From the path integral formalism, the entropy production in the non-equilibrium steady state (i.e., change in the Shannon entropy is zero) is expressed by
\begin{eqnarray}
\Sigma=\int\mathrm{d}s \dot{\mathcal{S}}_\mathrm{bath}(s)+\int\mathrm{d}s \dot{\mathcal{S}}_\mathrm{pump}(s)
\end{eqnarray}
where $\dot{\mathcal{S}}_\mathrm{bath}$ and $\dot{\mathcal{S}}_\mathrm{pump}$ are the entropy production rates in thermal bath and controller. The later has been referred to as ``entropy pumping" \cite{munakata2012entropy,munakata2014entropy}, which gives second-law like inequality including the influence of information extraction as
\begin{eqnarray}
\dot{\mathcal{S}}_\mathrm{bath} \geq -\dot{\mathcal{S}}_\mathrm{pump}.
\end{eqnarray}

To derive the actual expression of entropy production in our quadratic measurement feedback, we start from the Langevin equation in the laboratory frame with the displacement $X$ and momentum $P$ as follows:
\begin{eqnarray}
&\dot{X}=P,\\
&\dot{P}+A(X,P,t)=\mathcal{F}_\mathrm{th},
\end{eqnarray}
where $A(X,P,t)$ is the term of the equation of motion specified in Eq. (\ref{fullLang}).
From the Fokker-Planck equation, 
\begin{eqnarray}
\partial_t\mathcal{P}=&\mathcal{L}\mathcal{P}
\end{eqnarray}
with an operator
\begin{eqnarray}
\mathcal{L}\equiv \partial_P A(X,P,t)+k_BT \partial_P^2-\partial_X(P/m),
\end{eqnarray}
we achieve the forward transition probability in stochastic path from the initial condition $(X_0,P_0,t_0)$ as
\begin{eqnarray}
\mathcal{P}(X,P,t|X_0,P_0,t_0)=&\mathcal{B}\exp\left[-\frac{1}{4k_BT\Gamma}\int_{t_0}^t \mathrm{d}s \left(\dot{P}(s)+A(X(s),P(s),s)\right)^2\right]\nonumber\\
&\times\exp\left[\frac{1}{2}\int_{t_0}^t\mathrm{d}s\frac{\partial A(X,P,t)}{\partial P}\right],\label{forP}
\end{eqnarray}
where $\mathcal{B}$ is a constant. The first exponential term corresponds to the Onsager-Machlup function, and the second term is derived from the Ito formula \cite{imparato2006fluctuation}. Thus, the conjugate dynamics is given by
\begin{eqnarray}
\mathcal{P}^*(X_0,P_0,t|X,P,t_0)=&\mathcal{B}\exp\left[-\frac{1}{4k_BT\Gamma}\int_{t_0}^t \mathrm{d}s \left(\dot{P}(s)+A^\ast(X(s),P(s),s)\right)^2\right]\nonumber\\
&\times\exp\left[\frac{1}{2}\int_{t_0}^t\mathrm{d}s\left(\frac{\partial A(X,P,t)}{\partial P}\right)^\ast\right],\label{conP}
\end{eqnarray}
where $^\ast$ denotes a time-reversal operation. Because the entropy production is given by the ratio between Eqs. (\ref{forP}) and (\ref{conP}), the time-reversal parity of $A(X,P,t)$ is crucial. From Eq. (\ref{fullLang}), $A(X,P,t)$ is given by
\begin{eqnarray}
A(X,P,t)&=\Gamma P+\Omega ^2 X-G_0\Omega\mathrm{erfc}\left(\frac{\mathcal{M}(X,P)}{\sqrt{2}\sigma_M}\right)\cos2\Omega t X\nonumber\\
&=\Gamma P+\Omega ^2 X-G_0\Omega\left[1-\mathrm{erf}\left(\frac{\mathcal{M}(X,P)}{\sqrt{2}\sigma_M}\right)\right]\cos2\Omega t X,\nonumber\\
\end{eqnarray}
where the complementary error function is decomposed to a constant and a odd function (error function). This decomposition is crucial for calculating its conjugate dynamics as
\begin{eqnarray}
A^\ast(X,P,t)&=\Gamma P^\ast+\Omega ^2 X^\ast-G_0\Omega\left[1-t_\mathrm{P}\mathrm{erf}\left(\frac{\mathcal{M}^\ast(X,P)}{\sqrt{2}\sigma_M}\right)\right]\cos2\Omega t X^\ast,\nonumber\\
&=-\Gamma P+\Omega ^2 X-G_0\Omega\left[1-\mathrm{erf}\left(\frac{\mathcal{M}(X,P)}{\sqrt{2}\sigma_M}\right)\right]\cos2\Omega t X.\nonumber\\
\end{eqnarray}
Here, $t_p=\{-1,1\}$ is determined by the time-reversal parity of the target observable $\mathcal{M}(X,P)$, where $t_p=1$ ($-1$) when the $\mathcal{M}^\ast(X,P)=\mathcal{M}(X,P)$ [$\mathcal{M}^\ast(X,P)=-\mathcal{M}(X,P)$]. Thus, regardless of the time-reversal parity of the target observable, the feedback force is treated as a reversible force in the conjugate dynamics \cite{munakata2014entropy}.  By using this probability in the conjugate dynamics, 
\begin{eqnarray}
&\exp[\Sigma]=\frac{\mathcal{P}(X,P,t|X_0,P_0,t_0)}{\mathcal{P}^*(X_0,P_0,t|X,P,t_0)}\nonumber\\
&=\exp\left[-\frac{1}{k_BT}\int\mathrm{d}s P\circ \left(\dot{P}+\Omega^2 X-G_0\Omega\left[1-\mathrm{erf}\left(\frac{\mathcal{M}(X,P)}{\sqrt{2}\sigma_M}\right)\right]\cos2\Omega t X\right)\right]\nonumber\\
&\times \exp\left[\frac{2G_0\Omega}{\sqrt{2\pi}\sigma_M}\int\mathrm{d}s \exp\left[-\left(\frac{\mathcal{M}(X,P)}{\sqrt{2}\sigma_M}\right)^2\right]\frac{\partial \mathcal{M}(X,P)}{\partial P}\circ X \cos2\Omega t\right],\nonumber\\
\label{EP1}
\end{eqnarray}
where $\circ$ explicitly denotes Stratonovich integral. The first exponential term corresponds to the entropy production in the thermal bath, and the second exponential term corresponds to the entropy production in the controller. The entropy production rate, which is directly achieved by taking the time derivative in Eq. (\ref{EP1}), can be expressed as follows:
\begin{eqnarray}
\dot{\Sigma}&= \dot{\mathcal{S}}_\mathrm{bath}+\dot{\mathcal{S}}_\mathrm{pump},\\
\dot{\mathcal{S}}_\mathrm{bath}&=-\frac{1}{k_BT}P\circ \left(-\Gamma P+F_\mathrm{th}\right),\\
\dot{\mathcal{S}}_\mathrm{pump}&=-\frac{2G_0\Omega}{\sqrt{2\pi}\sigma_M}\exp\left[-\left(\frac{\mathcal{M}(X,P)}{\sqrt{2}\sigma_M}\right)^2\right]\frac{\partial \mathcal{M}(X,P)}{\partial P}\circ X \cos2\Omega t. \label{Spump_lab}
\end{eqnarray}
Because the target observable corresponds to $K_x$ which is given in the rotating frame, we perform the rotating wave approximation to linearize the transformation from the laboratory frame $(X,P)$ to the rotating frame ($q,p$) as
\begin{eqnarray}
X=&q\cos\Omega t+p\sin \Omega t,\\
P/\Omega\approx&-q\sin\Omega t+p\cos\Omega t.
\end{eqnarray}
Thus, the momentum derivative of the target observable in Eq. (\ref{Spump_lab}) is evaluated by
\begin{eqnarray}
\frac{\partial \mathcal{M}(X,P)}{\partial P}&\approx \frac{\partial q}{\partial P}\frac{\partial \mathcal{M}(q,p)}{\partial q}+ \frac{\partial p}{\partial P}\frac{\partial \mathcal{M}(q,p)}{\partial p}\nonumber\\
&=\frac{1}{2\Omega}(q\cos\Omega t-p\sin\Omega t).
\end{eqnarray}
By using the following approximation,
\begin{eqnarray}
&P^2\approx 2\Omega^2 K_z,\\
&(q\cos\Omega t-p\sin\Omega t)X\cos2\Omega t/(2\Omega)\approx K_z/2,
\end{eqnarray}
we obtain the expressions of stochastic average of entropy production,
\begin{eqnarray}
\left\langle \dot{\mathcal{S}}_\mathrm{bath}\right\rangle&\approx \frac{2\Omega^2\Gamma}{k_BT}\left(\langle K_z \rangle-\frac{k_BT}{2\Omega^2}\right),\label{Sheat}\\
\left\langle \dot{\mathcal{S}}_\mathrm{pump}\right\rangle&\approx \sqrt{\frac{1}{2\pi}} \frac{\tilde{G}_0\Gamma}{\sigma_M}\left\langle K_z\exp\left(-\frac{K_x^2}{2\sigma_M^2}\right)\right\rangle.\label{Spump}
\end{eqnarray}

\section{Entropy production in our protocol from probability currents}
Although the path integral formalism shown in Appendix C provides us a complete expression of entropy production with exact physical meaning, attempting to calculate it via the another simple formalism directly from the coarse-grained Fokker-Planck equation via probability currents \cite{tome2010entropy,spinney2012entropy,horowitz2014second} is worthwhile to confirm our formula in Eqs. (\ref{Sheat})-(\ref{Spump}). The Fokker-Planck equation is re-expressed by probability currents $J_z$ $(z=X,P)$ as
\begin{eqnarray}
\partial_t \mathcal{P}&=-\sum_{z} \partial_z J_z,\\
J_X&=-P\mathcal{P},\\
J_P&=\left(-\Gamma P-\Omega^2 X+G_0\Omega\left[1- \mathrm{erf}\left(\frac{\mathcal{M}(X,P)}{\sqrt{2}\sigma_M}\right)\right]\cos 2\Omega t X-k_BT\Gamma\partial_P\right)\mathcal{P}.\nonumber\\
\end{eqnarray}
Here, we split the momentum current $J_P$ into two in terms of the time-reversal parity, the same as discussed in Appendix C, in which the feedback force is regarded as a reversible force,
\begin{eqnarray}
J_P^\mathrm{rev}&=\left[-\Omega^2 X+G_0\Omega \left[1-\mathrm{erf}\left(\frac{\mathcal{M}(X,P)}{\sqrt{2}\sigma_M}\right)\right]\cos 2\Omega t X\right]\mathcal{P},\\
J_P^\mathrm{irr}&=\left[-\Gamma P-k_BT\Gamma \partial_P\right]\mathcal{P}.\label{flux1}
\end{eqnarray} 

Introducing Shannon entropy $\langle \mathcal{S}\rangle\equiv-\int\mathrm{d}X\mathrm{d}P\mathcal{P} \mathrm{ln}\mathcal{P}$ using a relationship $\partial_t \langle \mathcal{S}\rangle=-\int\mathrm{d}X\mathrm{d}P\left(\partial_t \mathcal{P}\right)\mathrm{ln}\mathcal{P}=-\int\mathrm{d}X\mathrm{d}P\left(\partial_t \mathcal{P}\right)(1+\mathrm{ln}\mathcal{P})$,
\begin{eqnarray}
\frac{\partial \langle \mathcal{S}\rangle }{\partial t}&=\int\mathrm{d}X\mathrm{d}P \left(\sum_z \partial_z J_z\right)\left(\mathrm{ln} \mathcal{P}+1\right)\\
&=-\int\mathrm{d}X\mathrm{d}P \sum_z \frac{J_z\partial_z \mathcal{P} }{\mathcal{P}}.
\end{eqnarray}
Here, the second equation is derived by using a partial integral and removing the boundary integral because the probability density function on the boundary is assumed to take zero. Note that
\begin{eqnarray}
\int\mathrm{d}X\mathrm{d}P \frac{J_x\partial_X \mathcal{P}}{\mathcal{P}}=-
\int\mathrm{d}X\mathrm{d}P \mathcal{P}\partial{X}\left(-P\right)=0.
\end{eqnarray}
Moreover, by using the relationship $\partial_P\mathcal{P}=-\frac{1}{k_BT\Gamma}\left(J_P^\mathrm{irr}+\Gamma P\right)$ from Eq. (\ref{flux1}), it reduces to
\begin{eqnarray}
\frac{\partial \langle \mathcal{S}\rangle }{\partial t}&=\frac{1}{k_BT\Gamma}\int\mathrm{d}X\mathrm{d}P \frac{1}{\mathcal{P}}\left[(J_P^\mathrm{irr})^2+\Gamma PJ_P^\mathrm{irr}\mathcal{P}\right]\nonumber\\
&-\int\mathrm{d}X\mathrm{d}P \frac{1}{\mathcal{P}} J_P^\mathrm{rev}\partial_P \mathcal{P}\nonumber\\
&=\langle \dot{\mathcal{S}}_\mathrm{tot}\rangle-\langle\dot{\mathcal{S}}_\mathrm{bath}\rangle-\langle\dot{\mathcal{S}}_\mathrm{pump}\rangle. \label{ineq1}
\end{eqnarray}
The first term in Eq. (\ref{ineq1}) corresponds to the non-negative entropy production,
\begin{eqnarray}
\langle \dot{\mathcal{S}}_\mathrm{tot} \rangle&\equiv\frac{1}{k_BT\Gamma}\int\mathrm{d}X\mathrm{d}P \frac{(J_P^\mathrm{irr})^2}{\mathcal{P}}\geq 0,
\end{eqnarray}
which obviously posses the second-law-like inequality,
\begin{equation}
\frac{\partial\langle\mathcal{S}\rangle}{\partial t}+\langle\dot{\mathcal{S}}_\mathrm{bath} \rangle+\langle\dot{\mathcal{S}}_\mathrm{pump} \rangle \geq 0. \label{ineq2}
\end{equation}
The second term in Eq. (\ref{ineq1}) corresponds to the entropy production rate due to the existence of irreversible currents. It can be expanded to
\begin{eqnarray}
\langle\dot{\mathcal{S}}_\mathrm{bath} \rangle&\equiv \frac{1}{k_BT\Gamma}\int\mathrm{d}X\mathrm{d}P \Gamma P\circ\left(\Gamma P+k_BT\Gamma \partial_P\right)\mathcal{P},\nonumber\\
&=\frac{1}{k_BT\Gamma}\Biggl(\Gamma^2 \langle P^2\rangle-k_BT\Gamma^2\Biggr),\nonumber\\
&\approx \frac{2\Omega^2\Gamma}{k_BT}\left(\langle K_z\rangle-\frac{k_BT}{2\Omega^2}\right)\label{entheat},
\end{eqnarray}
where $\tilde{G}_0\equiv G_0/\Gamma$ is notated. The approximation in Eq. (\ref{entheat}) is equivalent to that in Eqs. (\ref{Sheat}) and (\ref{Spump}). Apparently, we can confirm that Eq. (\ref{entheat}) completely corresponds to the entropy production in the thermal bath, Eq. (\ref{Sheat}), derived in the path integral formalism. 

The third term in Eq. (\ref{ineq1}) is regarded as the entropy production rate thanks to the presence of measurement and feedback, i.e., the entropy pumping rate, simplified as 
\begin{eqnarray}
\langle\dot{\mathcal{S}}_\mathrm{pump}\rangle&\int\mathrm{d}X\mathrm{d}P \frac{1}{\mathcal{P}} J_P^\mathrm{rev}\partial_P \mathcal{P}\nonumber,\\
&=G_0\Omega\cos2\Omega t\left\langle \partial_P\left[\mathrm{erf}\left(\frac{\mathcal{M}(X,P)}{\sqrt{2}\sigma_M}\right)\circ X\right] \right\rangle,\nonumber\\
&\approx \sqrt{\frac{1}{2\pi}}\frac{\tilde{G}_0\Gamma}{\sigma_M}\left\langle K_z\exp\left(-\frac{K_x^2}{2\sigma_M^2}\right)\right\rangle.
\end{eqnarray}
This expression is also equivalent to that in Eq. (\ref{Spump}) derived in the path integral formalism.

\section{Estimation of $\sigma_M$}
To estimate a noise deviation in the measurement $\sigma_M$, an observation is modelled by
\begin{equation}
K_x^M=a(K_x+\sigma_M\xi_M),
\end{equation}
where $a$ shows an arbitrary coefficient in measurement, and $\xi$ shows a Markovian noise with $\langle \xi(t)\xi(t')\rangle=\delta(t-t')$. Because the true value of $K_x$ can be approximated by $K_x^P$, which is the post-processed value, $\sigma_M$ can be determined by
\begin{eqnarray}
\sigma_M&=\sqrt{\frac{\langle (K_x^M)^2\rangle}{a^2}-\langle (K_x^P)^2\rangle}\\
a&=\frac{\langle K_x^M K_x^P\rangle}{\langle (K_x^P)^2\rangle}
\end{eqnarray}
from the experimental data without any driving (i.e., $\langle K_x^M \rangle=\langle K_x^P\rangle=0$). As a result, $\sigma_M$ is determined to be $0.52\pm 0.07$.

\section{Entropy production in continuous driving and random driving}
In the case of continuous driving (i.e., the feedback function becomes unity, $f(m)=1$), the expression of entropy production is straightforwardly derived because it only contains the contribution of the entropy production in the thermal bath, $\dot{\mathcal{S}}^\mathrm{C}_\mathrm{bath}$. Thus, $\langle\dot{\mathcal{S}}^\mathrm{C}_\mathrm{bath}\rangle$ is achieved with the same definition given in Eq. (\ref{Sheat}). In the same manner, the entropy production with the random protocol, in which the feedback function is given by $f(m)=\xi_R$ with the random integer $\xi_R\equiv\{0,1\}$, can be formulated by taking into account the contribution from the thermal bath $\langle\dot{\mathcal{S}}^\mathrm{R}_\mathrm{bath}\rangle$. Because entropy production is defined as the ratio between the forward and backward probability, we consider the minimum entropy production as that under the random switching. In other words, the force by the random switching is regarded as reversible in its conjugate dynamics, and as a result the total entropy production just consists of the entropy production in thermal bath \cite{munakata2012entropy,munakata2014entropy}. Consequently, this means $\langle\dot{\mathcal{S}}^\mathrm{R}_\mathrm{bath}\rangle$ can be calculated from Eq. (\ref{Sheat}). $\langle\dot{\mathcal{S}}^\mathrm{C}_\mathrm{bath}\rangle$ and $\langle\dot{\mathcal{S}}^\mathrm{R}_\mathrm{bath}\rangle$ can be analytically calculated by solving the following Langevin equation in the rotating frame:

\begin{eqnarray}
\langle\dot{K}_x\rangle=-\Gamma\langle K_x\rangle+\alpha G_0\langle K_z\rangle,\\
\langle\dot{K}_z\rangle=-\Gamma\langle K_z\rangle+\alpha G_0\langle K_x\rangle+\Gamma K_0,
\end{eqnarray}
where $K_0\equiv k_BT/2\Omega^2$, and $\alpha$ is a factor defined by $\alpha=1$ or $\alpha=1/2$ in case of continuous driving or random driving, respectively. The steady-state solutions are given by
\begin{eqnarray}
\langle K_z\rangle=\frac{K_0}{1-\alpha^2\tilde{G}_0^2},\hspace{20pt}\langle K_x\rangle=\frac{\alpha\tilde{G}_0K_0}{1-\alpha^2\tilde{G}_0^2}. \label{anaK}
\end{eqnarray}
By substituting them into Eq. (\ref{Sheat}), the entropy production rates under the random protocol are analytically expressed as follows:
\begin{eqnarray}
\langle\dot{\mathcal{S}}^\mathrm{C}_\mathrm{bath} \rangle=\frac{\Gamma\tilde{G}_0^2}{1-\tilde{G}_0^2},\label{Sc}\\
\langle\dot{\mathcal{S}}^\mathrm{R}_\mathrm{bath} \rangle=\frac{\Gamma\tilde{G}_0^2/4}{1-\tilde{G}_0^2/4}\label{Sran}
\end{eqnarray}
It is obvious that the entropy production in continuous driving (random driving) holds a divergence at $\tilde{G}_0=1$ ($\tilde{G}_0=2$). This divergence occurs because the thermal bath cannot absorb the heat from the parametric driving due to $G_0\geq \Gamma$ (or $\tilde{G}_0/2\geq \Gamma$). 

\section*{References}
\bibliographystyle{unsrt}

\end{document}